\documentclass[twocolumn,showpacs,prb]{revtex4}
\usepackage{graphicx}% Include figure files

\begin{document}

\preprint{}
\title{
Flux flow and pinning of the vortex sheet structure 
in a two-component superconductor
}

\author{Yasushi Matsunaga} 
%\email{matsunaga@mp.okayama-u.ac.jp}
\author{Masanori Ichioka} 
\email{oka@mp.okayama-u.ac.jp}
\author{Kazushige Machida}
%\email{machida@mp.okayama-u.ac.jp}
\affiliation{Department of Physics, Okayama University,
         Okayama 700-8530, Japan}

\date{\today}

\begin{abstract}
A simulation study using the time-dependent Ginzburg-Landau theory 
is performed for the vortex state in two-component superconductors, 
such as ${\rm PrOs_4Sb_{12}}$. 
We investigate the flux flow and the pinning of the vortex sheet structure.  
We find domain wall that traps half flux-quantum vortices  
and moves with the flux flow. 
In the pinning case, we observe an emitting process 
of a conventional vortex from the vortex sheet by combining 
a pair of half flux-quantum vortices. 
\end{abstract}


\maketitle 
%%%%%%%%%%%%%%%%%%%%%%%%%%%%%%%%%%%%%%%%%%%%%%%%%%%%%%
%\section{Introduction}
%\label{sec:introduction}

Some exotic unconventional superconductors have 
multiple superconducting phases, 
indicating a multi-component pair potential of the superconductivity. 
In these superconductors, we expect interesting behavior 
to occur through a combination of the multiple components. 

In the newly discovered superconductor ${\rm PrOs_4Sb_{12}}$, 
which is a heavy fermion compound with a filled skutterdite structure, 
thermal conductivity experiments have shown a multiple superconducting 
phase diagram with two phases, the A (in the higher $H$ region) 
and B (lower $H$) phases.\cite{Izawa}
Multi-component superconductivity is also suggested by 
the spontaneous moment in the superconducting phase 
reported in muon spin relaxation ($\mu$SR) experiments.~\cite{Aoki} 
As an another example, 
${\rm UPt_3}$ has a double transition of superconductivity 
and the superconducting phase is divided into three phases, 
indicating multi-component 
superconductivity.\cite{Adenwalla,Bruls,Dijk,Machida,Fujita}

Considering the two component case for simplicity, 
the pair potential in a $2 \times 2$ matrix form can be decomposed as  
$\hat\Delta({\bf r},{\bf k})
=\eta_1({\bf r})\hat\phi_1({\bf k})
+\eta_2({\bf r})\hat\phi_2({\bf k})$  
with the order parameter $\eta_m({\bf r})$, 
where ${\bf r}$ is the center of mass coordinate of the Cooper pair 
and $m=1,2$. 
The pairing function, depending on the relative momentum ${\bf k}$ 
of the pair, is given by 
$\hat\phi_m({\bf k})=i \hat\sigma_y\phi_m({\bf k})$ 
for the spin singlet pairing, and 
$\hat\phi_m({\bf k})=i \sum_{j=x,y,z}
d_{m,j}({\bf k})\hat\sigma_j \hat\sigma_y $ 
for the triplet pairing 
with Pauli matrices $\hat{\sigma}_x$, $\hat{\sigma}_y$, and $\hat{\sigma}_z$. 
In the double transition of superconductivity, 
after the first component $\eta_1$ appears at the first transition, 
the second component $\eta_2$ appears at the second 
transition.\cite{Goryo,Matsunaga}
We discuss in this paper the exotic vortex structure due to the combination of 
the two components below the second transition. 

Superfluidity of ${\rm ^3He}$, which has A and B phases, 
is also a typical example of multi-component pair potentials. 
Chiral $p$-wave superconductivity in ${\rm Sr_2RuO_4}$ should be 
treated as a two-component superconductor with  
$k_x+ i k_y$-wave and $k_x- i k_y$-wave components, 
or $k_x$-wave and $k_y$-wave components, 
when we consider the vortex,~\cite{Heeb,Takigawa,IchiokaP}
the domain wall and boundary states.\cite{Matsumoto} 

In multi-component superconductors, some superconducting states 
are degenerate in free energy. 
In a simple example of weak coupling singlet pairing  
with $\phi_1=\phi_{\rm A}$ and  $\phi_2=i\phi_{\rm B}$, 
two states 
$\Delta_\pm=\eta_1 \phi_{\rm A} \pm i \eta_2 \phi_{\rm B}$ 
are degenerate when $|\Delta_+|=|\Delta_-|$. 
Therefore, a domain structure may appear in a 
multi-component superconductor, i.e., some regions in a sample are 
a $\Delta_+$ domain and others are a $\Delta_-$ domain.  
Between the domains, domain walls appear as 
topological defects, which are not easily destroyed. 

When magnetic fields are applied to this domain structure, 
some of the vortices are trapped at the domain wall. 
The vortices at the domain wall form an exotic structure called a 
``vortex sheet'',\cite{Parts} where a conventional vortex splits 
into two vortices with half flux-quanta.\cite{Sigrist,BabaevL,BabaevB}
We confirmed the appearance of the vortex sheet 
by a simulation in a previous paper.\cite{Matsunaga} 
This interesting vortex sheet structure can be a clue 
to the presence of the domain wall, which would be clear evidence of 
unconventional multi-component superconductors. 

When a supercurrent flows in the vortex state, vortices flow 
in the direction transverse to the current, 
which is the origin of the flux flow resistivity. 
In order to stop the vortex motion and prevent flux flow resistivity, 
we need to introduce pining centers for vortices in the sample. 
The purpose of this paper is to investigate the dynamics of 
the vortex sheet structure by simulation of flux flow 
and pinning based on the time-dependent 
Ginzburg-Landau (TDGL) theory, 
and to examine how vortices and the domain 
wall of the vortex sheet structure move under the current flow 
in the presence of pinning centers. 
These dynamics may affect the flux flow resistivity 
or the magnetization process in multi-component superconductors. 
The TDGL theory is used as a phenomenological approach in our 
qualitative study, with the expectation that vortices move so as to approach 
the free energy minimum state. 
Contrary to a full-gap superconductor, 
since unconventional superconductors have low energy quasiparticle states 
available for the dissipation process 
within the superconducting gap due to the gap node, 
we expect that the TDGL theory can be qualitatively applied 
without considering the extreme case of the dirty limit. 

We start from the two-component Ginzburg-Landau (GL) theory 
that was used in our previous study on the double transition 
of ${\rm PrOs_4Sb_{12}}$. 
The GL free energy density in the dimensionless form is 
written as~\cite{Matsunaga}
\begin{eqnarray} && 
\tilde{f} 
=-\left(\nu({\bf r})-\frac{T}{T_{{\rm c}}}\right)|\eta_1|^2 
 -\left(\frac{T_{{\rm c}2}}{T_{{\rm c}}}\nu({\bf r})
       -\frac{T}{T_{{\rm c}}}\right)|\eta_2|^2 
\nonumber \\ && 
+ \eta_1^\ast (q_x^2+q_y^2) \eta_1 
+C_{22x} \eta_2^\ast q_x^2 \eta_2 
+C_{22y} \eta_2^\ast q_y^2 \eta_2 
\nonumber \\ && \hspace{0.5cm} 
+C_{12x} \eta_1^\ast q_x^2 \eta_2 +C_{12x}^\ast \eta_2^\ast q_x^2 \eta_1  
\nonumber \\ && \hspace{0.5cm}  
+C_{12y} \eta_1^\ast q_y^2 \eta_2 +C_{12y}^\ast \eta_2^\ast q_y^2 \eta_1  
\nonumber \\ && 
+\frac{1}{2} \{ |\eta_1|^4 +C_2 |\eta_2|^4 +4C_3
 |\eta_1|^2  |\eta_2|^2  
\nonumber \\ && \hspace{0.5cm}  
+C_4^\ast \eta_1^{\ast 2} \eta_2^2 
+C_4      \eta_2^{\ast 2} \eta_1^2 \},  \qquad 
\label{eq:fe}
\end{eqnarray} 
where ${\bf q}=(\hbar/ i)\nabla+(2\pi/\phi_0){\bf A}$ 
with flux-quantum $\phi_0$ and vector potential ${\bf A}$, 
and we write $C_{22x}\equiv (1-\epsilon)/(X\sqrt{1-\epsilon^2})$ and 
$C_{22y}\equiv (1+\epsilon)/(X\sqrt{1-\epsilon^2})$. 
Inside the pinning center region, 
where the superconductivity is suppressed, 
we set $\nu({\bf r})=0$.\cite{KatoC} 
In other regions, $\nu({\bf r})=1$. 
The GL theory covers both singlet and triplet pairings 
and we can study the vortex states without specifying 
the pairing function $\hat{\phi}({\bf k})$. 
The coefficients in Eq. (\ref{eq:fe}) are related to the information 
of the pairing symmetry and the Fermi surface anisotropy 
as described in Ref. \onlinecite{Matsunaga}. 
However, since this information has not been established yet, 
we treat the coefficients as arbitrary parameters. 

We consider a phase below the second 
transition, where both $\eta_1$ and $\eta_2$ appear. 
From Eq. (\ref{eq:fe}), we recognize that the relative phases of 
$\eta_1$ and $\eta_2$ should be $(\alpha+\pi)/2$ or $(\alpha - \pi)/2$ 
in the free energy minimum state in a zero field, 
where $\alpha$ is given by $C_4=|C_4| e^{i \alpha}$. 
These two states correspond to $\Delta_+$ and $\Delta_-$ 
noted in the introduction. 

\begin{figure} [htb] 
\includegraphics[width=4.5cm]{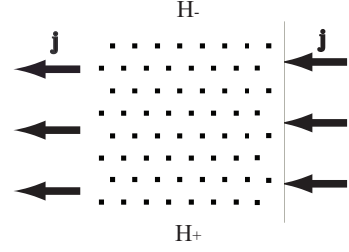} 
\caption{
Configuration of the simulation. 
The current density $j$ flows in at the right-hand side boundary and out at the left-hand side boundary
of $W \times W$ square superconductors. 
External fields $H_+$ and $H_-$ are applied outside the lower and upper boundaries, respectively. 
For the pinning case, we introduce pinning centers 
where the superconductivity is suppressed, 
represented by small solid squares.   
} 
\label{fig:f1}
\end{figure}

For simulations of the vortex dynamics we use 
the TDGL equation coupled with Maxwell equations:~\cite{KatoC,Kato,MMachida} 
\begin{eqnarray} && 
\frac{\partial}{\partial t}\eta_1 
= -\frac{1}{12} \frac{\partial \tilde{f}}{\partial \eta_1^\ast}, 
\qquad 
\frac{\partial}{\partial t}\eta_2 
= -\frac{1}{12} \frac{\partial \tilde{f}}{\partial \eta_2^\ast}, 
\label{eq:TDGL}
\\ && 
\frac{\partial}{\partial t}{\bf A} =  \tilde{\bf j}_{\rm s} 
-\kappa^2 \nabla\times{\bf B}, 
\qquad 
{\bf B}=\nabla\times{\bf A}. 
\label{eq:Maxwell} 
\end{eqnarray} 
The supercurrent $\tilde{\bf j}_{\rm s}
=(\tilde{j}_{{\rm s},x},\tilde{j}_{{\rm s},y})
\propto(\partial \tilde{f}/\partial A_x, \partial \tilde{f}/\partial A_y)$ 
is given by 
\begin{eqnarray} &&
\tilde{j}_{{\rm s},x}
={\rm Re} [ 
         \eta_1^\ast (q_x \eta_1) 
+C_{22x} \eta_2^\ast (q_x \eta_2) 
\nonumber \\ && \hspace{1cm} 
+C_{12x} \eta_1^\ast (q_x \eta_2) 
+C_{12x}^\ast \eta_2^\ast (q_x \eta_1)  ],  \qquad 
\label{eq:jsx}
\\ && 
\tilde{j}_{{\rm s},y}
={\rm Re}[  \eta_1^\ast (q_y \eta_1 )
+C_{22y} \eta_2^\ast (q_y \eta_2 )
\nonumber \\ && \hspace{1cm} 
+C_{12y} \eta_1^\ast (q_y \eta_2) 
+C_{12y}^\ast \eta_2^\ast (q_y \eta_1)  ]. 
\label{eq:jsy}
\end{eqnarray} 
The length, field, and time are, respectively, scaled by 
the coherence length $\xi_0$, $H_{{\rm c}2,0}=\phi_0/2\pi \xi_0^2$, and 
$t_0=4 \pi \xi_0^2 \kappa^2 \sigma / c^2$ with the 
normal state conductivity $\sigma$.\cite{KatoC,Kato,MMachida} 
However, we here scale $\eta_m$ by $\eta_0$ instead of 
$\eta_0(T)=\eta_0(1-T/T_{{\rm c}})^{1/2}$. 
$\eta_0$ is a uniform solution of $\eta_1$ when 
$\eta_2=0$, $\nu({\bf r})=1$, and $T=0$. 
The calculations are performed in a two-dimensional square area 
with each side of length of $W$ and a current flow $j$, 
as schematically shown in Fig. \ref{fig:f1}. 
Outside the open boundary we set $\eta_1=\eta_2=0$, 
and at the lower and upper boundaries we set $B({\bf r})=H_+$ 
and $H_-$, respectively,
where $H_\pm=H_0 \pm Wj/2$ for an applied field $H_0$.\cite{MMachida} 

%%%%%%%%%%%%%%%%%%%%%%
%%%%%%%%%%%%%%%%%%%%%%

We report the case when the gradient coupling 
is absent ($C_{12x}=C_{12y}=0$).\cite{Ichioka} 
We set $T_{{\rm c}2}/T_{\rm c}=0.9$, $\kappa=4$, 
$C_2=1$, $C_3=-C_4=0.2$, $\epsilon=0$, and $X=0.5$. 
If $C_4$ is negative (i.e., $\alpha=\pi$), then  
$\phi_1=\phi_{\rm A}$ and  $\phi_2=i\phi_{\rm B}$, 
if the pairing is singlet. 
We note that our results for the vortex sheet do not significantly 
depend on the selection of the coefficients in Eq. (\ref{eq:fe}), 
because the appearance of the domain wall only results 
from two states with the relative phase $(\alpha\pm\pi)/2$ being 
degenerate in free energy. 
We report results for $T=0.1T_{\rm c}$, 
$H=0.2 H_{{\rm c}2,0}$, $j=2.6 \times 10^{-3}$, and $W=57 \xi_0$ 
as a typical example. 
In this simulation, we start from a uniform initial state 
with relative phase $0$, and switch on $H_0$ and $j$ at $t=0$.  

In some cases of the penetration process of vortices, 
vortex sheet structures are created at the boundary 
and move toward the inside.\cite{Matsunaga} 
On the other hand, the vortex sheet can be produced if we apply 
a magnetic field after preparing the domain wall at a zero field. 
In this case, 
some penetrating vortices are trapped at the domain wall 
and these trapped vortices become a vortex sheet by 
splitting into half flux-quantum vortices.  

%%%%%%%%%%%%%%%%%%%%%%
\begin{figure} [htb] 

\vspace{0.5cm} 
\includegraphics[width=6.5cm]{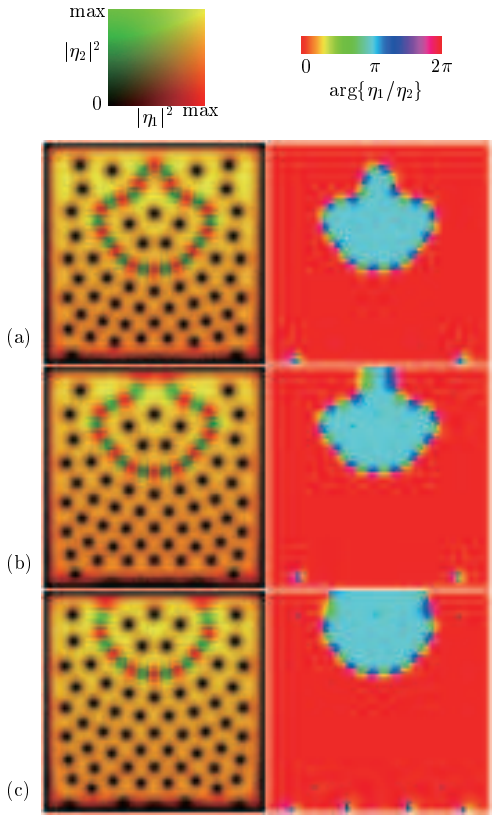} 

\caption{
Time evolution of the vortex state in the free flux flow case 
without pinning centers. 
Snapshots of the total $W \times W$ area are presented for 
(a) $t/54t_0=17$, (b) 22, and (c) 27.  
The left panels show color density plots of 
$|\eta_1({\bf r})|$ and $|\eta_2({\bf r})|$. 
The black region is the core of the conventional vortex. 
The green and red regions are the cores of the $\eta_1$ 
and $\eta_2$ vortices, respectively, 
along the vortex sheet. 
The right panels show the relative phase 
${\arg}\{ \eta_1({\bf r})/\eta_2({\bf r})\}$. 
The red and blue regions are the domains of relative phase 
$0$ and $\pi$, respectively. 
} 
\label{fig:f2}
\end{figure} 
%%%%%%%%%%%%%%%%%%%%%%

First, we study the free flux flow case without pinning centers, 
as shown in Fig. \ref{fig:f2}.
The left panels show a color map of the spatial distribution of 
$|\eta_1({\bf r})|$ and $|\eta_2({\bf r})|$. 
The black circle region shows a conventional vortex with a flux-quantum, 
where $|\eta_1({\bf r})|$ and $|\eta_2({\bf r})|$ 
share the same vortex core. 
The green and red circle regions shows the $\eta_1$ 
and $\eta_2$ vortices, respectively, 
where only $|\eta_1({\bf r})|$ or $|\eta_2({\bf r})|$, respectively, 
has a vortex core 
and the other, $|\eta_2({\bf r})|$ or $|\eta_1({\bf r})|$, does not. 
These green and red vortex cores are located alternatively along a loop, 
forming a vortex sheet. 
The area including both an $\eta_1$ vortex and an $\eta_2$ vortex 
is equal to that of a conventional vortex, indicating that each of 
the two vortices $\eta_1$ and 
$\eta_2$ has half flux-quanta. 
The right panels shows the relative phase 
${\arg}\{ \eta_1({\bf r})/\eta_2({\bf r}) \}$, 
where we recognize that the vortex sheet is along the 
domain wall between the domains of relative phase $\pi$ (blue region) and 
relative phase 0 (red region). 
The relative phase has a phase winding $2\pi$ or $-2\pi$ around 
the $\eta_1$ or $\eta_2$ vortex, respectively. 
In this free flux flow case, a conventional vortex flows upward, i.e., 
in a perpendicular direction to the current. 
The vortex sheet structure also simply flows 
with the same flow as conventional vortices. 
That is, the domain wall moves with the trapping half flux-quantum 
vortices under current flow.  
The half flux-quantum vortices rarely escape 
from the domain wall of the vortex sheet in this free flux flow case.  

%%%%%%%%%%%%%%%%%%%%%%
%%%%%%%%%%%%%%%%%%%%%%
\begin{figure} [htb] 

\vspace{0.5cm} 
\includegraphics[width=6.5cm]{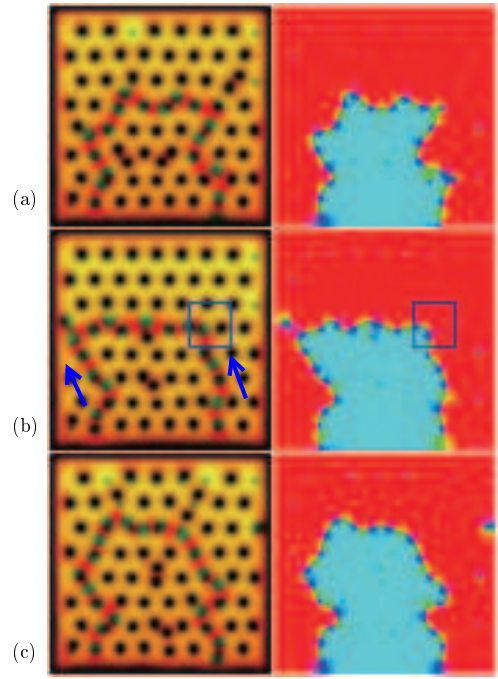} 

\caption{
Time evolution of the vortex state in the presence of pinning centers. 
Snapshots are presented for (a) $t/54t_0=196$, (b) 276, and (c) 356.  
The left panels show color density plots of 
$|\eta_1({\bf r})|$ and $|\eta_2({\bf r})|$. 
The right panels show the relative phase 
${\arg}\{ \eta_1({\bf r})/\eta_2({\bf r})\}$. 
} 
\label{fig:f3}
\end{figure} 
%%%%%%%%%%%%%%%%%%%%%%

To study dynamical processes involving the vortex sheet and 
conventional vortices in a complicated manner, 
we introduce pinning centers to prevent the flux flow. 
Figure \ref{fig:f3} shows the time evolution of the flux flow 
states in the presence of pinning,  where 
pinning centers with a $1.4 \xi_0 \times 1.4 \xi_0$ square area are 
introduced periodically, as presented in Fig. \ref{fig:f1}.  
Vortices move slowly, repeatedly  
being trapped at pinning centers and escaping from them. 
The line of the vortex sheet also moves slowly and meanders 
due to the trapping of $\eta_1$ and $\eta_2$ vortices 
by the pinning centers. 
Compared with the slow motion of a conventional vortex and the line 
of the vortex sheet, $\eta_1$  and $\eta_2$ vortices easily flow 
along the domain wall of the vortex sheet,  
as indicated by arrows in Fig. \ref{fig:f3}(b). 
It seems that the barrier for escaping from a pinning center 
is lower along the line of the vortex sheet, 
because the vortex sheet is incommensurate with the pinning centers 
and the intervortex distance is short in this direction. 
Therefore, in the pinning case the domain wall forms a ``channel'' 
through which the vortex can flow. 
When the domain wall is connected to the sample boundary, 
vortices penetrate into the sample through the domain wall channel.
These half flux-quantum vortices, quickly moving through the channel, 
are emitted from the vortex sheet at the front corner of the 
slowly moving vortex sheet line. 
At the emitting point, pairs of half flux-quantum vortices 
combine to become a conventional vortex 
and the created conventional vortex is released from the domain wall 
of the vortex sheet. 

%%%%%%%%%%%%%%%%%%%%%%
%
\begin{figure*} [tbh] 
%\begin{figure} [htb] 

%\vspace{0.5cm} 
\includegraphics[width=9.0cm]{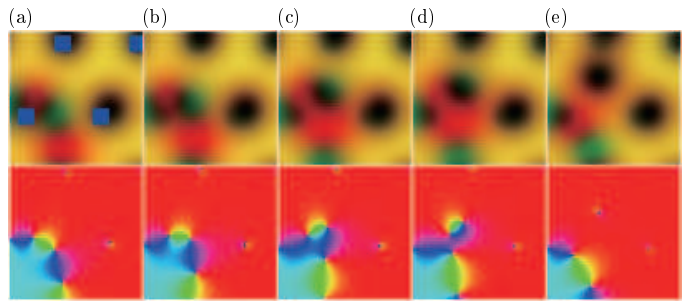} 

\caption{
Time evolution of the creation of a conventional vortex from 
half flux-quantum vortices in the vortex sheet. 
Snapshots at (a) $t/54t_0=276$, (b) 284, (c) 290, (d) 292, and (e) 296 
are presented within the enclosed area shown in Fig. 3(b).  
The upper panels show color density plots of 
$|\eta_1({\bf r})|$ and $|\eta_2({\bf r})|$. 
The positions of the pinning centers are represented by 
purple squares in (a). 
The lower panels show the relative phase 
${\arg}\{ \eta_1({\bf r})/\eta_2({\bf r})\}$. 
The winding center between the yellow and purple regions is 
a vortex center of $\eta_1$ or $\eta_2$ vortices. 
} 
\label{fig:f4}
\end{figure*} 
%\end{figure} 
%%%%%%%%%%%%%%%%%%%%%%

Figure \ref{fig:f4} shows the time evolution of 
this emitting process in an enclosed region in Fig. \ref{fig:f3}(b). 
In the upper panel in Fig.  \ref{fig:f4}(a) 
we see that vortices are trapped around pinning centers, 
represented by purple square regions. 
In the lower panels in Fig.  \ref{fig:f4}, 
we find vortex centers of $\eta_1$ or $\eta_2$ vortices as 
winding centers of the relative phase (between the yellow and purple regions). 
In Fig.  \ref{fig:f4}(a), 
$\eta_1$ and $\eta_2$ vortices flow upward along the vortex sheet line. 
However, the vortex sheet cannot easily move, 
because some of the $\eta_1$ and $\eta_2$ vortices in the vortex 
sheet are trapped by pinning centers. 
Therefore, near the front corner of the domain wall shown in 
Fig.  \ref{fig:f4}, the intervortex distance of  $\eta_1$ and $\eta_2$ vortices 
becomes short and the curvature of the domain wall becomes sharp
[(a)$\rightarrow$(b)$\rightarrow$(c)]. 
This deformation of the domain wall shape helps the creation of the 
conventional vortex. 
After the core region of two $\eta_2$ vortices (the red regions in the upper panels) 
overlap, 
one of the $\eta_2$ vortices forms a pair with a neighbor 
$\eta_1$ vortex [(c)$\rightarrow$(d)]. 
This $\eta_1$ and $\eta_2$ vortex pair combines to become  
a conventional vortex (the black region) and leaves the domain wall [(e)].
We also observe the inverse process where 
a conventional vortex is trapped to the vortex sheet and changes to 
a pair of $\eta_1$ and $\eta_2$ vortices in the pinning case simulation. 
This event is rare, but occasionally observed. 

In summary, 
we have performed a TDGL simulation for the flux flow and the pinning in 
a two-component superconductor,  
and investigated the dynamics of the domain wall and half flux-quantum 
vortices of the vortex sheet structure. 
The domain wall moves with the flux flow both in the free flux flow 
case and the pinning case. 
We succeeded in observing the creation process wherein 
a pair of half flux-quantum vortices is changed to a conventional 
vortex and released from the vortex sheet. 
Relating these phenomena, the vortex sheet structure may 
contribute to the flux flow resistivity  
or the magnetization process in multi-component superconductors, 
such as in ${\rm PrOs_4Sb_{12}}$, ${\rm UPt_3}$, and ${\rm Sr_2RuO_4}$.  
A quantitative estimate of these interesting contributions 
is left for future studies. 

%
%%%%%%%%%%%%%%%%%%%%%%%%%%%%%%%%%%%%%%%%%55

%%%% references %%%%%%%%%%%%%%%%%%%%%%%%%%%%%%%%%%%%%%%%%%%%%%%%%%%%

%\end{references}
%%%%%%%%%%%%%%%%%%%%%%%%%%%%%%%%%%%%%%%%%%%%%%%%%%%%%%%

\newpage
\begin{thebibliography}{99}
%\begin{references}

\bibitem{Izawa} 
K. Izawa, Y. Nakajima, J. Goryo, Y. Matsuda, S. Osaki, H. Sugawara, 
H. Sato, P. Thalmeier, and K. Maki,  
Phys. Rev. Lett. {\bf 90}, 117001 (2003). 

\bibitem{Aoki} 
Y. Aoki, A. Tsuchiya, T. Kanayama, S.R. Saha, H. Sugawara, H. Sato, 
W. Higemoto, A. Koda, K. Ohishi, K. Nishiyama, and R. Kadono, 
Phys. Rev. Lett. {\bf 91}, 067003 (2003)

\bibitem{Adenwalla} 
S. Adenwalla, S.W. Lin, Q.Z. Ran, Z. Zhao, J.B. Ketterson, 
J.A. Sauls, L. Taillefer, D.G. Hinks, M. Levy, and B.K. Sarma, 
Phys. Rev. Lett. {\bf 65}, 2298 (1990). 

\bibitem{Bruls}
G. Bruls, D. Weber, B. Wolf, P. Thalmeier, B. L\"{u}thi, 
A. de Visser, and A. Menovsky, 
Phys. Rev. Lett. {\bf 65}, 2294 (1990).

\bibitem{Dijk}
N.H. van Dijk, A. de Visser, J.J.M. Franse, S. Holtmeier, 
L. Taillefer, and J. Flouquet, 
Phys. Rev. B {\bf 48}, 1299 (1993).

\bibitem{Machida}  
K. Machida, M. Ozaki and T. Ohmi, 
J. Phys. Soc. Jpn. {\bf 58}, 4116 (1989). 

\bibitem{Fujita}  
T. Fujita, W. Aoyama, K. Machida, and T. Ohmi, 
J. Phys. Soc. Jpn. {\bf 63}, 247 (1994). 

\bibitem{Goryo}
J. Goryo, 
Phys. Rev. B {\bf 67}, 184511 (2003). 

\bibitem{Matsunaga} 
Y. Matsunaga, M. Ichioka, and K. Machida, 
Phys. Rev. Lett. {\bf 92}, 157001 (2004)

\bibitem{Heeb}
R. Heeb and D.F. Agterberg, 
Phys. Rev. B {\bf 59}, 7076 (1999).

\bibitem{Takigawa}
M. Takigawa, M. Ichioka, K. Machida, and M. Sigrist, 
Phys. Rev. B {\bf 65}, 014508 (2002). 

\bibitem{IchiokaP} 
M. Ichioka and K. Machida, 
Phys. Rev. B {\bf 65}, 224517 (2002).

\bibitem{Matsumoto}
M. Matsumoto and M. Sigrist, 
J. Phys. Soc. Jpn. {\bf 68}, 994 (1999). 

\bibitem{Parts}
\"{U}. Parts, E.V. Thuneberg, G.E. Volovik, J.H. Koivuniemi, 
V.M.H. Ruutu, M. Heinil\"{a}, J.M. Karim\"{a}ki, and M. Krusius, 
Phys. Rev. Lett. {\bf 72}, 3839 (1994). 

\bibitem{Sigrist} 
M. Sigrist and D.F. Agterberg, 
Prog. Theor. Phys. {\bf 102}, 965 (1999). 

\bibitem{BabaevL} 
E. Babaev, 
Phys. Rev. Lett. {\bf 89}, 067001 (2002). 

\bibitem{BabaevB} 
E. Babaev, L.D. Faddeev, and A.J. Niemi, 
Phys. Rev. B {\bf 65}, 100512 (2002). 

\bibitem{KatoC}
R. Kato, Y. Enomoto, and S. Maekawa, 
Physica C {\bf 227}, 387 (1994). 

\bibitem{Kato}
R. Kato, Y. Enomoto, and S. Maekawa, 
Phys. Rev. B {\bf 47}, 8016 (1993). 

\bibitem{MMachida}
M. Machida and H. Kaburaki, 
Phys. Rev. Lett. {\bf 71}, 3206 (1993). 

\bibitem{Ichioka}
M. Ichioka, N. Nakai, and K. Machida, 
J. Phys. Soc. Jpn. {\bf 72}, 1322 (2003).

\end{thebibliography}
\end{document}